\documentstyle[sprocl,epsfig]{article}       % for nijmegen
\def\Nbar{\bar N}               % only, or it won't work!
\def\nc{\bar n_c}
\def\as{\alpha_s}
\def\pt{p\kern -.2pt\lower 4pt\hbox{\tiny T}}    %works?

\def\p0{P_0(\Delta y)}
\def\Dy{\Delta y}

\def\kt{k_{\perp}}

\def\NF{{\mathcal N}_{\kern -1.9pt f}}
\def\NC{{\mathcal N}_{\kern -1.7pt c}}

\def\dyg{{\delta y_g}}
\def\cP{{\mathcal P}}
\def\cQ{{\mathcal Q}}
\def\HL{{H_j^L}}

\begin{document}
\thispagestyle{empty}
\begin{center}
\hfill LU TP 96-26\\
\hfill September 23rd, 1996\\
\vspace{2.4cm}

{\bf MULTIPLICITY DISTRIBUTIONS AND FLUCTUATIONS IN THE DISCRETE QCD MODEL}
\vspace{0.8cm}

        R. Ugoccioni \footnote[1]{E-mail: roberto@thep.lu.se}\\
{\it Department of Theoretical Physics, University of Lund,\\
   S\"olvegatan 14A, S-22362 Lund, Sweden}
\vspace{2cm}

ABSTRACT\\
\end{center}

\begin{quote}\noindent Results are presented from calculations carried
out in the discrete QCD model,
concerning multiplicity distributions in rapidity intervals 
and fluctuations.\end{quote}

\begin{center}
\vspace{2.5cm}
Talk presented at the\\ 7th International Workshop ``Correlations and
Fluctuation''\\ (Nijmegen, The Netherlands, June 30 -- July 6, 1996)
\end{center}
\newpage

\title{MULTIPLICITY DISTRIBUTIONS AND FLUCTUATIONS IN THE
DISCRETE QCD MODEL}

\author{R. Ugoccioni}
\address{Department of Theoretical Physics, University of Lund,\\
   S\"olvegatan 14A, S-22362 Lund, Sweden}
\maketitle

\abstracts{Results are presented from calculations carried
out in the discrete QCD model,
concerning multiplicity distributions in rapidity intervals 
and fluctuations.}

\section{Introduction}

One of the long standing problems in multiparticle dynamics is
an integrated description of multiplicity distributions and
correlations. In particular, from the
point of view of perturbative QCD, these observables are not
satisfactorily described: this is particularly
evident in restricted regions of
phase space, where the most interesting experimental results
are.\cite{VietriRU}
The discrete approximation to QCD~\cite{discreteQCD} 
allows to perform
exclusive and inclusive calculations at parton level;
in this talk,
a few results are described concerning Multiplicity Distributions (MD's)
and fluctuations.

\section{A short description of discrete QCD}

\begin{figure}[t]
\begin{center}
\mbox{\epsfig{file=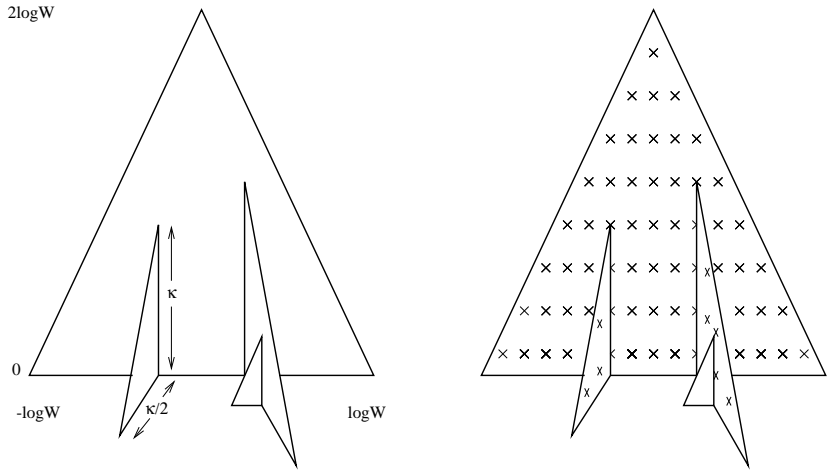,height=6.2cm}}
\end{center}
\caption[Phase space]{{\textbf{left)}} Phase space in the Dipole Cascade Model.
The dimensions shown are $y$ and $\kappa = \log\kt^2$, respectively
rapidity and log of transverse momentum, taken in each dipole's c.m.\ frame.
{\textbf{right)}} Phase space
in the discrete QCD model. Emission of ``effective gluons''
can only take place at the points marked with a cross.
}
\label{fig:triangle}
\end{figure}

Here I will present only the fundamental ideas behind this
model; for a detailed discussion, follow the pointers
in the bibliography.\cite{discreteQCD}
Consider an event in which a quark-antiquark pair is produced:
the cross section for the emission of gluons by this pair
is, in the Double Log Approximation (DLA), given by the 
``colour dipole radiation'' formula:\cite{QCDcoherence}
\begin{equation}
  dP \approx \frac{C \as}{\pi} \frac{d \kt^2}{\kt^2} dy
						\label{eq:dipole} 
\end{equation}
where $C$ is a constant depending on the colour charges of the partons
at the end points of the dipole, and $y$ and $\kt$ are the rapidity
and transverse momentum of the emitted gluon, respectively. Notice
that the azimuthal direction is here integrated over, but it is
of course possible to take it fully into account.
The phase space limits for the gluon emission are given 
approximately by
\begin{equation}
  |y| < \log(W/\kt)				\label{eq:triangle}
\end{equation}
where $W$ is the dipole mass (which in the case of the original
$q\bar q$ dipole equals the c.m.\ energy);
the rapidity range available for the first gluon emission is thus
$\Dy = \log(W^2/\kt^2)$. After the first emission, the three partons
form two independent dipoles: they will in turn emit gluons according to
Eq.~\ref{eq:dipole} (in the c.m.\ frame of each dipole)
and the cascade develops.\cite{Dipole}
This is illustrated in Figure \ref{fig:triangle} (left part)
where it is also shown that, because the gluons carry colour charges,
the phase space actually increases with each emission: a new double
faced triangular region is added for each new gluon emitted; 
notice that according to Eq.~\ref{eq:triangle} the baseline
of each face added is half the height of the same face.
The Modified Leading Log Approximation (MLLA) changes this picture
in that it provides
a region of phase space, close to the end points of the dipole, 
where gluon emission is suppressed.\cite{pQCD}
Thus the effective rapidity range becomes smaller:
for example, for a $gg$ dipole one has for large $W$:
\begin{equation}
  \Dy = \log\frac{W^2}{\kt^2} - \dyg		\label{eq:mlla}
\end{equation}
It can be argued,\cite{discreteQCD,Gustafson} from the form of the beta term 
in the Callan-Symanzik equation, that $\dyg$ equals $11/6$.
Dipole splitting with ordering in $\kt$ can also
be shown to be equivalent to the angular ordering prescription
in standard parton cascades.\cite{discreteQCD,pQCD}

In the discrete QCD model,\cite{discreteQCD} 
the idea expressed in Eq.~\ref{eq:mlla} is carried to its
extreme consequences: it is assumed that if two gluons are
closer than $\dyg$ they will effectively be re-absorbed into a single
gluon. One needs then only consider ``effective gluons'': the 
``rapidity'' dimension of phase
space is thus partitioned in discrete steps of size $\dyg$.
However, this is true also for every triangle ``added'' to the
original phase space, which means that discrete steps must be taken
also in the vertical dimension, i.e., in $\kappa \equiv \log \kt^2$: 
phase space is
thus partitioned in rectangular cells of size $\dyg \times 2\dyg$.
In a complementary way, one can say that emission of effective gluons
takes place only at a discrete lattice of points, $\dyg$ units apart
horizontally and $2 \dyg$ units apart vertically
(see Figure~\ref{fig:triangle}). 
From the very structure of the cascade described above,
it follows that each ``column'' of cells is independent of the others:
such a column will be called a {\em tree} because of the
possibility of emitting further triangles (``branches'').
The probability of emitting a gluon with transverse momentum
$\kappa$ in such a cell can be obtained from Eq.~\ref{eq:dipole}
using the leading order expression for the running coupling
constant $\as$ and the appropriate colour factor for the $g\to gg$
vertex:
\begin{equation}
  dP = \frac{1}{\dyg} \frac{d\kappa}{\kappa} \dyg = \frac{d\kappa}{\kappa}
\end{equation}
Normalizing with a Sudakov factor, the distribution in $\kappa$
for an effective gluon at a given rapidity is uniform:
\begin{equation}
  d {\mathrm{Prob}} = \exp\left\{ -\int_\kappa^{\kappa_{\mathrm{max}}}
	dP \right\} dP = \frac{d\kappa}{\kappa_{\mathrm{max}}}
						\label{eq:dprob}
\end{equation}
where $\kappa_{\mathrm{max}}$ corresponds to the maximum $\kt$ available
at that rapidity.
Each triangle emerging from the original one can then be described
by the position and length of its baseline (both as 
integer multiples of $\dyg$):
an events is a collection
of independent trees of different height, the number of trees
being given by $L = 2\log (W/\dyg)$. The maximum height $\HL$ 
of each tree (numbering, e.g., the trees from left to right)  
depends only on the tree's position $j$ and on the event ``baseline
length'' $L$.
A tree of height $H$ corresponds to an emitted gluon of
transverse momentum $\kappa = H \cdot 2 \dyg$. A tree of height $0$
corresponds to the emission of no gluons; notice that this introduces
a natural cut-off for cascade evolution. Notice also that
only a finite number of gluons can be emitted. These
points are important in characterizing discrete QCD.

In order to calculate the multiplicity distribution,
one can now define an {\em $N$-tree} as a tree with height between 
$0$ and $N-1$ (in units of $\dyg$) where $N$ is an integer.
It is convenient to define also a {\em true $N$-tree} as a tree
of height $N-1$. Because of Eq.~\ref{eq:dprob}, an $N$-tree is
obtained by summing true $H$-trees ($H=1,\ldots,N$) with equal 
probability $1/N$. This is easily expressed in terms of
the generating functions for the MD's in an  $N$-tree, $\cP_N(z)$,
and in a true $N$-tree, $\cQ_N(z)$:
\begin{equation}
  \cP_N(z) = \frac{1}{N} \sum_{H=1}^{N} \cQ_H(z)	\label{eq:one}
\end{equation}
It will also be recognized that a true $(N+1)$-tree can be obtained from a
true $N$-tree by attaching two $N$-trees to it (one on each side)
\begin{equation}
  \cQ_{N+1}(z) = \cQ_N(z) \left[ \cP_N(z) \right]^2	\label{eq:two}
\end{equation}
Using the obvious initial conditions:
\begin{equation}
 \cP_1(z) = \cQ_1(z) = 1 ; \qquad \cQ_2(z) = z ; \qquad
	\cP_2(z) = \frac{1+z}{2}			\label{eq:initcond}
\end{equation}
Eq.s~\ref{eq:one} and \ref{eq:two} can be solved,\cite{discreteQCD}
thus giving the MD in a tree.

\section{Calculations in rapidity intervals}

In order to approach the calculation in rapidity intervals,
one should understand that the size of $\dyg$ does not
imply that there is no structure at distances smaller than $11/6$.
In fact, $\dyg$ correspond to a separation
in rapidity along the original $q\bar q$ axis (which one can identify
to a good approximation with the thrust axis of the event)
only for trees which `grow' from the original baseline, while inside
each branch, $\dyg$ is a distance along the local `jet' direction. 
Thus one can say that gluons belonging to the same tree are all
localized in rapidity inside one unit of $\dyg$.
That this is
actually a good approximation has been checked with a Monte Carlo
implementation of the discrete QCD model. In this framework one can then
calculate observables integrated over intervals
of rapidity whose width is a multiple of $\dyg$, i.e., which contain
a whole number of trees. So one recognizes that the MD
in any of these rapidity interval is given by $\cP_{\HL}(z)$.
Adjacent trees are independent, and so are adjacent rapidity intervals
in this approximation; thus the MD generating function in an interval
$\Dy$ is given by 
\begin{equation}
  \cP_{\Dy}(z)\; = \prod_{j \in \Dy} \cP_{\HL}(z)
\end{equation}
where the product is over all trees inside the interval $\Dy$.

\begin{figure}[t]
\begin{center}
\mbox{\epsfig{file=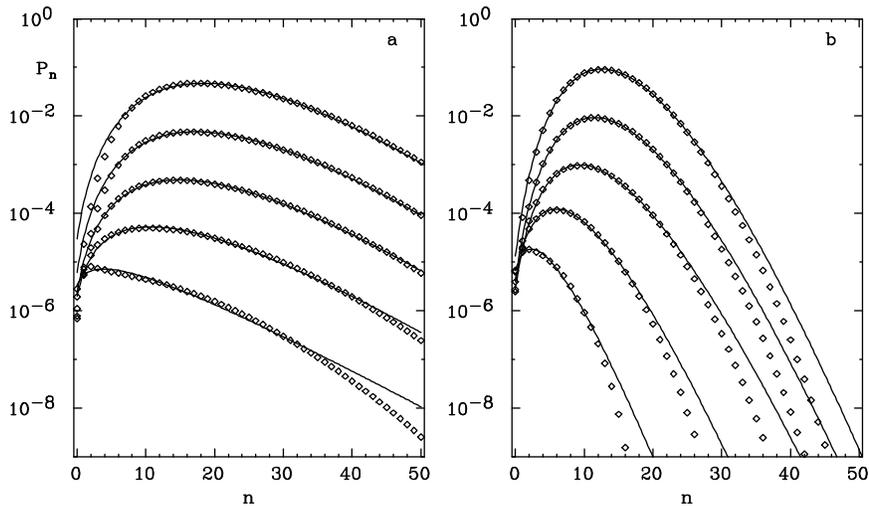,%
bbllx=23pt,bblly=78pt,bburx=550pt,bbury=388pt,width=11.5cm}}
\end{center}
\caption[MD in intervals]{{\textbf{a)}}
Multiplicity distributions in central intervals
of rapidity of size (from bottom to top) 2, 4, 6, 8 and 10 
(in units of $\dyg$) for standard events with baseline length $L=12$. 
Each distribution is divided by 10 with respect
to one above it in order to make the figure legible.
{\textbf{b)}} Same as in a), but for ``2-jet like'' events.}
\label{fig:md}
\end{figure}

Figure \ref{fig:md}, on the left, shows the resulting MD in 
different central rapidity intervals: the solid
line shows a negative binomial distribution (NBD), phenomenologically
widely used for describing the data.\cite{VietriRU}
The same figure shows on the right the MD for what I will call
in the following ``2-jet like'' events, obtained by cutting
away the tip of the phase space triangle,
i.e., requiring a cut-off
on the {\em maximum} $\kt$ of an emitted gluon.
Notice that in the standard events, for small intervals,
the NBD behaves w.r.t. the points in a way similar
to the way it behaves w.r.t to experimental data on hadronic MD,
going first below the data, than above and then below again.\cite{DEL:2}
In the 2-jet like events, on the other hand, 
the NBD is closer to the points than in the standard events:
one is reminded of the experimental result that a NBD is found
to describe the hadronic MD in 2-jet events (defined via a jet-finder
algorithm) better than in the full sample.\cite{DEL:4}

In the literature, the NBD is often explained in terms of
clan structure:\cite{AGLVH:2,FaroAG} 
independent emission of clans, according to a Poisson distribution,
followed by particle emission from each clan in a cascade way (logarithmic
distribution).
In the framework of discrete
QCD, it is tempting to notice that each tree which comes from the
baseline of the original triangle is independent from the others;
a tree which produces no partons is not ``observed'', so the number
of trees in a rapidity interval is variable.
One can therefore call trees of height larger than $0$
{\em non-empty trees}; by definition these contain at least one parton. 
\begin{figure}[!t]
\begin{center}
\mbox{\epsfig{file=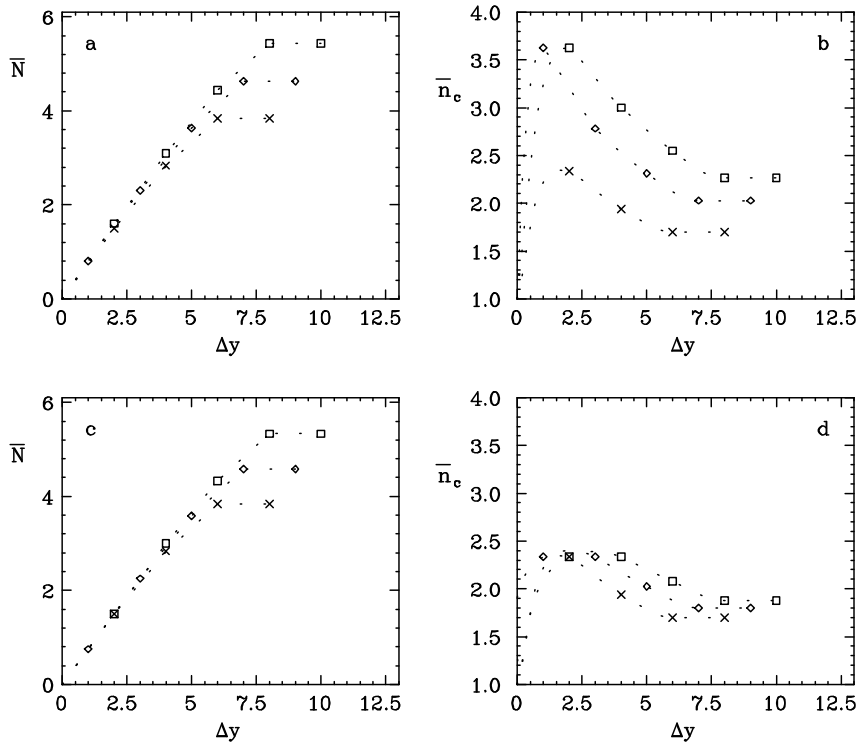,bbllx=61pt,%
bblly=266pt,bbury=672pt,bburx=528pt,width=11.5cm}}
\end{center}
\caption[clans]{{\textbf{a)}} Average number of non-empty trees in
rapidity intervals vs size of the interval (in units of $\dyg$),
for three sizes of the event: $L=8$ (crosses), $L=9$ (diamonds)
and $L=10$ (squares).
{\textbf{b)}} Average multiplicity in an average non-empty tree
vs size of the rapidity interval for the said event sizes.
{\textbf{c)}} As in a), but for 2-jet like events.
{\textbf{d)}} As in b), but for 2-jet like events.}
\label{fig:clans}
\end{figure}

Consider now an event with $L$ trees:
\begin{equation}
  x_j \equiv \cP_{\HL}(0) = \frac{1}{\HL}
\end{equation}
is the probability that the $j$-th tree is empty. The MD in
the number of non-empty trees $n$ is given by the product of
the probabilities that $L-n$ trees are empty and $n$ are not:
\begin{equation}
  P_n^{(L)} = \sum_{\mathrm{perm}} (1-x_1)\ldots(1-x_n) x_{n+1}\ldots x_L
	\qquad n=0,\ldots,L
\end{equation}
where the sum is over all permutations of $x$'s.
It can be shown that $P_n^{(L)}$ satisfies the recurrence relation
\begin{equation}
  P_n^{(L)} = \sum_{j=1}^n \frac{(-1)^{j+1}}{n} a_j P_{n-j}^{(L)}
\end{equation}
where
\begin{equation}
  a_r = \sum_{i=1}^L \left( \frac{1}{x_i} - 1\right)^r \qquad{\mathrm and}
	\qquad P_0^{(L)} = \prod_{i=1}^L x_i
\end{equation}
From these relations $P_n^{(L)}$ can be calculated;
it is easy to show that one obtains a binomial distribution
when $x_j$ is independent of $j$, and a Poisson distribution
when $a_r = \delta_{1,r}$: the former is incompatible with
Eq.~\ref{eq:one}, but the latter is a good approximation
when $x_j$ is not very small.
\begin{figure}[t]
\begin{center}
\mbox{\epsfig{file=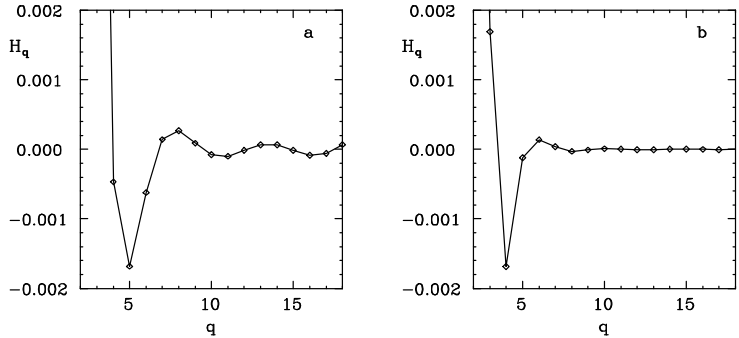,bbllx=36pt,bblly=480pt,%
bburx=576pt,bbury=709pt,width=11.0cm}}
\end{center}
\caption[Hq moments]{{\textbf{a)}} Ratio of factorial cumulant moments to
factorial moments, $H_q$, vs the order of the moments, $q$,
for standard events of baseline length $L=10$.
{\textbf{b)}} Same as in a) but for 2-jet like events.}
\label{fig:hq}
\end{figure}
Figure \ref{fig:clans}
shows the result for the average number of non-empty trees $\Nbar(\Dy)$ and
for the average multiplicity within an average non-empty tree $\nc(\Dy)$
(defined as the ratio of the average multiplicity to $\Nbar(\Dy)$)
for standard events. The noticeable features for the average number
of non-empty trees are an independence
from the size $L$ of the event in small intervals, and a linearity
with the size of the interval at fixed $L$. Both these characteristics
are seen in experimental data with the statistical definition of
clans.\cite{DEL:2,Elba}

Finally, in Figure \ref{fig:hq} the results of
a calculation in full phase space of the ratio $H_q$
of factorial cumulant moments to factorial moments
are shown: these where shown to be sensitive to the structures
in the MD,\cite{NijmegenAG,FaroRU} and in particular to the radiation
of hard gluons in the early stages of the partonic evolution.
The figure shows that oscillations in sign of the ratio $H_q$
appear in the discrete QCD model, and when
examined in the 2-jet like sample these oscillations diminish
strongly in amplitude, in accordance with the behaviour of
data.\cite{NijmegenAG}

\section*{Acknowledgments}
I would like to thank Bo Andersson and Jim Samuelson for
very useful discussions. A warm thank goes also to the organizers
of this excellent meeting.

%\vspace{-0.2cm}
\section*{References}
%auto-ignore
% this file is read in by the main file via \input.
%%%% References file: dqcd.ref
%%%% Creator: REFLATEX.PL 1.0
%%%% Usage: \input{dqcd.ref} where references should go.
%%%%

\end{document}